# KEY QUESTIONS FOR MODELLING COVID-19 EXIT STRATEGIES

## AUTHORS

Robin N. Thompson[1,2,3,*], T. Déirdre Hollingsworth[4], Valerie Isham[5], Daniel Arribas-Bel[6,7], Ben Ashby[8], Tom Britton[9], Peter Challenor[10], Lauren H. K. Chappell[11], Hannah Clapham[12], Nik J. Cunniffe[13], A. Philip Dawid[14], Christl A. Donnelly[15,16], Rosalind M. Eggo[3], Sebastian Funk[3], Nigel Gilbert[17], Julia R. Gog[18], Paul Glendinning[19], William S. Hart[1], Hans Heesterbeek[20], Thomas House[21,22], Matt Keeling[23], István Z. Kiss[24], Mirjam E. Kretzschmar[25], Alun L. Lloyd[26], Emma S. McBryde[27], James M. McCaw[28], Joel C. Miller[29], Trevelyan J. McKinley[30], Martina Morris[31], Philip D. O'Neill[32], Carl A. B. Pearson[3,33], Kris V. Parag[16], Lorenzo Pellis[19], Juliet R. C. Pulliam[33], Joshua V. Ross[34], Michael J. Tildesley[23], Gianpaolo Scalia Tomba[35], Bernard W. Silverman[15,36], Claudio J. Struchiner[37], Pieter Trapman[9], Cerian R. Webb[13], Denis Mollison[38], Olivier Restif[39]

## AFFILIATIONS

[1]Mathematical Institute, University of Oxford, Woodstock Road, OX2 6GG Oxford, UK
[2]Christ Church, University of Oxford, St Aldates, Oxford OX1 1DP, UK
[3]Department of Infectious Disease Epidemiology, London School of Hygiene and Tropical Medicine, Keppel Street, London WC1E 7HT, UK
[4]Big Data Institute, University of Oxford, Old Road Campus, Oxford OX3 7LF, UK
[5]Department of Statistical Science, University College London, Gower Street, London WC1E 6BT, UK
[6]School of Environmental Sciences, University of Liverpool, Brownlow Street, Liverpool L3 5DA, UK
[7]The Alan Turing Institute, British Library, 96 Euston Road, London NW1 2DB, UK
[8]Department of Mathematical Sciences, University of Bath, North Road, Bath BA2 7AY, UK
[9]Department of Mathematics, Stockholm University, Kräftriket, 106 91 Stockholm, Sweden
[10]College of Engineering, Mathematical and Physical Sciences, University of Exeter, Exeter EX4 4QE, UK
[11]Department of Plant Sciences, University of Oxford, South Parks Road, Oxford, OX1 3RB, UK
[12]Saw Swee Hock School of Public Health, National University of Singapore, 12 Science Drive, Singapore 117549, Singapore
[13]Department of Plant Sciences, University of Cambridge, Downing Street, Cambridge, CB2 3EA, UK
[14]Statistical Laboratory, University of Cambridge, Wilberforce Road, Cambridge, CB3 0WB, UK
[15]Department of Statistics, University of Oxford, St Giles', Oxford OX1 3LB, UK
[16]Department of Infectious Disease Epidemiology, Imperial College, Norfolk Place, London W2 1PG, UK
[17]Department of Sociology, University of Surrey, Stag Hill, Guildford GU2 7XH, UK
[18]Centre for Mathematical Sciences, University of Cambridge, Wilberforce Road, Cambridge CB3 0WA, UK

*Correspondence to: robin.thompson@chch.ox.ac.uk

[19]Department of Mathematics, University of Manchester, Oxford Road, Manchester M13 9PL, UK
[20]Department of Population Health Sciences, Utrecht University, Yalelaan, 3584 CL Utrecht, The Netherlands
[21]IBM Research, The Hartree Centre, Daresbury, Warrington WA4 4AD, UK
[22]Mathematics Institute, University of Warwick, Gibbet Hill Campus, Coventry CV4 7AL, UK
[23]Zeeman Institute for Systems Biology and Infectious Disease Epidemiology Research, School of Life Sciences and Mathematics Institute, University of Warwick, Gibbet Hill Road, Coventry CV4 7AL, UK
[24]School of Mathematical and Physical Sciences, University of Sussex, Falmer, Brighton BN1 9QH, UK
[25]Julius Center for Health Sciences and Primary Care, University Medical Center Utrecht, Utrecht University, Heidelberglaan 100, 3584CX Utrecht, The Netherlands
[26]Department of Mathematics, North Carolina State University, Stinson Drive, Raleigh, NC 27607, USA
[27]Australian Institute of Tropical Health and Medicine, James Cook University, Townsville, Queensland 4811, Australia
[28]School of Mathematics and Statistics, University of Melbourne, Carlton, Victoria 3010, Australia
[29]Department of Mathematics and Statistics, La Trobe University, Bundoora, Victoria 3086, Australia
[30]College of Medicine and Health, University of Exeter, Barrack Road, Exeter EX2 5DW, UK
[31]Department of Sociology, University of Washington, Savery Hall, Seattle, Washington 98195, USA
[32]School of Mathematical Sciences, University of Nottingham, University Park, Nottingham NG7 2RD, UK
[33]South African DSI-NRF Centre of Excellence in Epidemiological Modelling and Analysis (SACEMA), Stellenbosch University, Jonkershoek Road, Stellenbosch 7600, South Africa
[34]School of Mathematical Sciences, University of Adelaide, South Australia 5005, Australia
[35]Department of Mathematics, University of Rome Tor Vergata, 00133 Rome, Italy
[36]Rights Lab, University of Nottingham, Highfield House, Nottingham NG7 2RD, UK
[37]Escola de Matemática Aplicada, Fundação Getúlio Vargas, Praia de Botafogo, 190 Rio de Janeiro, Brazil
[38]Department of Actuarial Mathematics and Statistics, Heriot-Watt University, Edinburgh EH14 4AS, UK
[39]Department of Veterinary Medicine, University of Cambridge, Madingley Road, Cambridge CB3 0ES, UK

# ABSTRACT

Combinations of intense non-pharmaceutical interventions ("lockdowns") were introduced in countries worldwide to reduce SARS-CoV-2 transmission. Many governments have begun to implement lockdown exit strategies that allow restrictions to be relaxed while attempting to control the risk of a surge in cases. Mathematical modelling has played a central role in guiding interventions, but the challenge of designing optimal exit strategies in the face of ongoing transmission is unprecedented. Here, we report discussions from the Isaac Newton Institute "Models for an exit strategy" workshop (11-15 May 2020). A diverse community of modellers who are providing evidence to governments worldwide were asked to identify the main questions that, if answered, will allow for more accurate predictions of the effects of different exit strategies. Based on these questions, we propose a roadmap to facilitate the development of reliable models to guide exit strategies. The roadmap requires a global collaborative effort from the scientific community and policy-makers, and is made up of three parts: i) improve estimation of key epidemiological parameters; ii) understand sources of heterogeneity in populations; iii) focus on requirements for data collection, particularly in Low-to-Middle-Income countries. This will provide important information for planning exit strategies that balance socio-economic benefits with public health.

# INTRODUCTION

As of 21 July 2020, the coronavirus disease 2019 (COVID-19) pandemic has been responsible for more than 14 million reported cases worldwide, including over 613,000 deaths. Mathematical modelling is playing an important role in guiding interventions to reduce the spread of Severe Acute Respiratory Syndrome Coronavirus 2 (SARS-CoV-2). Although the impact of the virus has varied significantly across the world, and different countries have taken different approaches to counter the pandemic, many national governments introduced packages of intense non-pharmaceutical interventions (NPIs), informally known as "lockdowns". Although the socio-economic costs (e.g. job losses and potential long-term mental health effects) are yet to be assessed fully, public health measures have led to substantial reductions in transmission [1–3]. Data from countries such as Sweden and Japan, where epidemics peaked without strict lockdowns being introduced, will be useful for comparing different approaches and conducting retrospective cost-benefit analyses.

As case numbers have either stabilised or declined in many countries, attention has now turned to the development of strategies that allow restrictions to be lifted [4,5] in order to alleviate the economic, social and other health costs of lockdowns. However, in countries with active transmission still occurring, daily disease incidence could increase again quickly, while countries that have suppressed community transmission successfully face the risk of transmission reestablishing due to reintroductions. In the absence of a vaccine or sufficient herd immunity to reduce transmission substantially, COVID-19 exit strategies pose unprecedented challenges to policy-makers and the scientific community. Given our limited knowledge of this virus, and the fact that entire packages of interventions were introduced in quick succession in many countries as case numbers increased, it is challenging to estimate the effects of removing individual measures directly and modelling remains of paramount importance.

Here, we report discussions from the "Models for an exit strategy" workshop (11-15 May 2020) that took place online as part of the Isaac Newton Institute's "Infectious Dynamics of Pandemics" programme. The Isaac Newton Institute in Cambridge is the UK's national research institute for mathematics, and many distinguished scientists (including nine Nobel laureates and 27 Fields Medallists) have attended programmes there. We outline progress to date and open questions in modelling exit strategies that arose during discussions at the workshop. Most participants were working actively on COVID-19 at the time of the workshop, often with the aim of providing evidence to governments, public health authorities and the general public to support the pandemic response. After four months of intense model development and data analysis, the workshop gave participants a chance to take stock and openly share their views of the main challenges they are facing. A range of countries were represented, providing a unique forum to discuss the different epidemic dynamics and policies around the world. Although the main focus was on epidemiological models, the interplay with other disciplines formed an integral part of the discussion. The purpose of this article is twofold: to highlight key knowledge gaps hindering current predictions and projections, and to provide a roadmap for modellers and other scientists wishing to make useful contributions to the development of solutions.

Given that SARS-CoV-2 is a newly discovered virus, the evidence base is changing rapidly. This makes it challenging to conduct a systematic review of the literature. For that reason, we asked the large group of researchers at the workshop for their expert opinions on the most important open questions, and relevant literature, that will enable exit strategies to be planned with more precision. By inviting contributions from representatives of different countries and areas of expertise (including social scientists, immunologists, infectious disease outbreak modellers and others), and discussing the expert views raised at the workshop in detail, we sought to reduce geographic and disciplinary biases. All evidence is summarised here in a policy-neutral manner.

The questions in this article have been grouped as follows. First, we discuss outstanding questions for modelling exit strategies that are related to key epidemiological quantities, such as the reproduction number and herd immunity fraction. We then identify different sources of heterogeneity underlying SARS-CoV-2 transmission and control, and consider how differences between hosts and populations across the world should be included in models. Finally, we discuss current challenges relating to data requirements, focussing on the data that are needed to resolve current knowledge gaps and how uncertainty in modelling outputs can be communicated to policy-makers and the wider public. In each case, we outline the most relevant issues, summarise expert knowledge and propose specific steps towards the development of evidence-based exit strategies. This leads to the development of a roadmap for future research (Fig 1) made up of three key steps: i) improve estimation of epidemiological parameters using outbreak data from different countries; ii) understand heterogeneities within and between populations that affect virus transmission and interventions; iii) focus on data needs, particularly data collection and methods for planning exit strategies in Low-to-Middle-Income countries (LMICs) where data are often lacking. This roadmap is not a linear process: improved understanding of each aspect of the proposed research will help to inform other requirements. For example, a clearer understanding of the model resolution required for accurate forecasting (Section 2.1) will inform the data that need to be collected (Section 3), and vice versa. If this roadmap can be followed, it will be possible for policy-makers to predict the effects of different potential exit strategies with increased precision. This is of clear benefit to global health, allowing exit strategies to be chosen that allow interventions to be relaxed while limiting the risk of substantial further transmission.

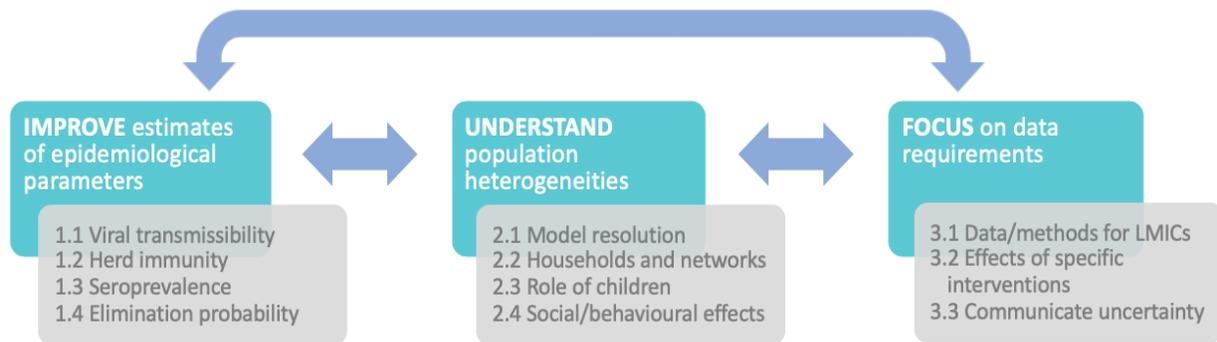

Figure 1. Roadmap of research to facilitate the development of reliable models to guide exit strategies. Three key steps are required: i) improve estimates of epidemiological parameters (such as the reproduction number and herd immunity fraction) using data from different countries (Sections 1.1-1.4); ii) understand heterogeneities within and between populations that affect virus transmission and interventions (Sections 2.1-2.4); iii) focus on data requirements for predicting the effects of individual interventions, particularly – but not exclusively – in data limited settings such as LMICs (Sections 3.1-3.3). Work in these areas must be conducted concurrently, since feedback will arise from the results of the proposed research that will be useful for shaping next steps across the different topics.

# 1 KEY EPIDEMIOLOGICAL QUANTITIES

## 1.1 HOW CAN VIRAL TRANSMISSIBILITY BE ASSESSED MORE ACCURATELY?

The time-dependent reproduction number, $R(t)$ or $R_t$, has emerged as the main quantity used to assess the transmissibility of SARS-CoV-2 in real time [6–10]. Within a given population with active virus transmission, the value of $R(t)$ represents the expected number of secondary cases generated by someone infected at time $t$. If this quantity is and remains below one, then an ongoing outbreak will eventually fade out. Although easy to understand intuitively, estimating *R(t)* from case reports (as opposed to, for example, observing *R(t)* in known or inferred transmission trees [11]) requires the use of mathematical models. As factors such as contact rates between infectious and susceptible individuals change during an outbreak in response to public health advice or movement restrictions, the value of $R(t)$ has been found to respond rapidly. For example, across the UK, countrywide and regional estimates of $R(t)$ dropped from approximately 2.5-4 in mid-March [7,12] to below one after lockdown was introduced [12,13]. One of the criteria in the UK and elsewhere for relaxing the lockdown was for the reproduction number to decrease to "manageable levels" [14]. Monitoring $R(t)$, as well as case numbers, as individual components of the lockdown are relaxed is critical for understanding whether or not the outbreak remains under control [15].

Several mathematical and statistical methods for estimating temporal changes in the reproduction number have been proposed in the last 20 years. Two popular approaches are the Wallinga–Teunis method [16] and the Cori method [17,18]. These methods use case notification data along with an estimate of the serial interval distribution (the times between successive cases in a transmission chain) to infer the value of $R(t)$. Other approaches exist (e.g. based on compartmental epidemiological models [19]), including those that can be used alongside different data (e.g. time series of deaths [7,12,20] or phylogenetic data [21–24]).

Despite this extensive theoretical framework, practical challenges remain when dealing with real-time reporting. In particular, reproduction number estimates often rely on case notification data and so are subject to reporting delays between case onset and being recorded. Available data therefore do not include up-to-date knowledge of current numbers of infections, an issue that can be addressed using "nowcasting" models [8,12,25]. The serial interval represents the period between symptom onset times in a transmission chain, rather than between the times at which cases are recorded. Time series of symptom onset dates, or even infection dates (to be used with the estimates of the generation interval when inferring $R(t)$), can be estimated from case notification data using latent variable methods [8,26] or deconvolution methods such as the Richardson-Lucy deconvolution technique [27,28]. The Richardson-Lucy approach has previously been applied to infer incidence curves from time series of deaths [29]. These methods, as well as others that account for reporting delays [30], provide useful avenues to improve the practical estimation of $R(t)$ given incomplete data. Furthermore, changes in testing practice (or capacity to conduct tests) lead to temporal changes in case numbers that cannot be distinguished easily from changes in transmission. Understanding how accurately and how quickly changes in $R(t)$ can be inferred in real-time given these challenges is a crucial question.

A more immediate way to assess temporal changes in the reproduction number that does not require nowcasting is by observing people's transmission-relevant behaviour directly, e.g. through contact surveys or mobility data [31]. These methods do, however, come with their own limitations: because these surveys do not usually collect data on infections, care must be taken in using them to understand and predict ongoing changes in transmission.

Other outstanding challenges in assessing variations in $R(t)$ include the need to understand that methods tend to be inaccurate when case numbers are low, and the requirement to account for temporal changes in the serial interval or generation time distribution of the disease [32]. Indeed, in periods when there are few cases (such as in the "tail" of an epidemic – Section 1.4), there is little information with which to assess virus transmissibility. Methods for estimating $R(t)$ that are based on the assumption that transmissibility is constant within fixed time periods can be applied with windows of long duration (thereby including more case notification data with which to estimate $R(t)$) [33,34]. However, this comes at the cost of a loss of sensitivity to temporal variations in transmissibility. Consequently, when case numbers are low, the methods described above for tracking transmission-relevant behaviour directly are particularly important. In those scenarios, the "transmission potential" might be more important than realised transmission [35].

The effect of population heterogeneity on reproduction number estimates also requires further investigation, as current estimates of $R(t)$ tend to be calculated for whole populations (e.g. countries or regions). Understanding the characteristics of constituent groups contributing to this value is important to target interventions efficiently with limited resources [36,37]. For this, data on infections within and between different subpopulations (e.g. infections in care homes and in the wider population) are needed. As well as between subpopulations, it is also necessary to ensure that estimates of $R(t)$ account for heterogeneity in transmission between different infectious hosts. Such heterogeneity alters the effectiveness of different control measures, and therefore the predicted disease dynamics when individual interventions are relaxed. For a range of diseases, a rule of thumb that around 20% of infected individuals are the sources of 80% of infections has been proposed [37,38]. This is supported by recent evidence for COVID-19, which suggests significant individual-level variation in SARS-CoV-2 transmission [39] with some transmission events leading to large numbers of new infections.

Finally, it is well documented that presymptomatic individuals (and, to a lesser extent, asymptomatic infected individuals - i.e. those who never develop symptoms) can transmit SARS-CoV-2 [40,41]. For that reason, negative serial intervals may occur when an infected host displays COVID-19 symptoms before the person who infected them [42,43]. Although methods for estimating $R(t)$ with negative serial intervals exist [43,44], the inclusion of presymptomatic or asymptomatic transmission in publicly available software for estimating $R(t)$ should be a priority. Increasing the accuracy of estimates of $R(t)$ in the ways described here, as well as supplementing these estimates with other assessments of transmissibility (e.g. estimates of growth rates of case numbers [45]), is of clear importance. As lockdowns are relaxed, this will permit a fast determination of whether or not removed interventions are leading to a surge in cases.

## 1.2 WHAT IS THE HERD IMMUNITY THRESHOLD AND WHEN MIGHT WE REACH IT?

Herd immunity refers to the accumulation of sufficient immunity in a population through infection and/or vaccination to prevent further substantial outbreaks. It is a major factor in determining exit strategies, but data are still very limited. Dynamically, the threshold at which herd immunity is achieved is the point at which $R(t)$ (Section 1.1) falls below one for an otherwise uncontrolled epidemic, resulting in a negative epidemic growth rate. However, reaching the herd immunity threshold does not mean that the epidemic is over or that there is no risk of further infections. Great care must therefore be taken in communicating this concept to the public, to ensure continued adherence to public health measures such as social distancing. Crucially, whether immunity is gained naturally through infection or through random or targeted vaccination affects the herd immunity threshold, and the threshold depends critically on the immunological characteristics of the pathogen. Since SARS-CoV-2 is a new virus, its immunological characteristics - notably the duration and extent to which prior infection confers protection against future infection, and how these vary across the population - are currently unknown [46]. Lockdown measures have heavily impacted contact structure and hence the accumulation of immunity in the population, and are likely to have led to significant heterogeneity in acquired immunity (e.g. by age, location, workplace). Knowing the extent and distribution of immunity in the population will help guide optimal exit strategies that have only a limited risk of a resurgence in infections.

As interventions are lifted, whether or not $R(t)$ remains below one depends on the current level of immunity in the population as well as the specific exit strategy followed. A simple illustration is to treat the current reproduction number, $R(t)$, as a deflation of the original (basic) reproduction number ($R_0$, which is assumed to be greater than one):

$$R(t) = (1 - i(t))(1 - p(t))R_0,$$

in which $i(t)$ is the immunity level in the community at time $t$ and $p(t)$ is the overall reduction factor from the control measures that are in place. If $i(t) > 1 - \frac{1}{R_0}$, then $R(t)$ remains below one even when all interventions are lifted: herd immunity is achieved. However, recent results [47,48] show that, for heterogeneous populations, herd immunity occurs at a lower immunity level than $1 - \frac{1}{R_0}$. The threshold $1 - \frac{1}{R_0}$ assumes random vaccination, with immunity distributed uniformly in the community. When immunity is obtained from disease exposure, the more socially active individuals in the population are over-represented in cases from the early stages of the epidemic. As a result, the virus preferentially infects individuals with higher numbers of contacts, thereby acting like a well-targeted vaccine. This reduces the herd immunity threshold. However, the extent to which heterogeneity in behaviour lowers the herd immunity threshold for COVID-19 is currently unknown.

We highlight three key challenges for determining the herd immunity threshold for COVID-19, and hence for understanding the impact of implementing or lifting control measures in different populations. First, most of the quantities for calculating the herd immunity threshold are not known precisely and require careful investigation. For example, determining the immunity level in a community is far from trivial for a number of reasons: antibody tests may have variable quality in terms of sensitivity and specificity; it is currently unclear whether or not individuals with mild or no symptoms acquire immunity or test seropositive; the duration of immunity is unknown. Second, estimation of $R_0$, despite receiving significant attention at the start of the pandemic, still needs to be refined within and between countries as issues with early case reports come to light and are addressed. Third, as discussed in Section 2, SARS-CoV-2 does not spread uniformly through populations [49]. An improved understanding of the main transmission routes, and which communities are most influential, will help to determine how much lower disease induced herd immunity is compared to the classical herd immunity threshold $1 - \frac{1}{R_0}$.

To summarise, it is of paramount importance to obtain more accurate estimates of the current immunity levels in different countries and regions, and to understand more clearly how population heterogeneity affects virus transmission and the accumulation of immunity.

## 1.3 CAN SEROPREVALENCE SURVEYS PROVIDE INSIGHT INTO HERD IMMUNITY AND TRANSMISSION DYNAMICS?

Information about how many people are currently infected and have been infected in the past are key inputs to formulate exit strategies, monitor the progression of epidemics and identify social and demographic sources of transmission heterogeneities. Seroprevalence surveys provide a simple and direct way to estimate the fraction of the population that has been exposed to the virus but has not been detected by regular surveillance mechanisms [50]. Given the possibility of mild or asymptomatic infections, which are not typically captured in data describing laboratory-confirmed cases, seroprevalence surveys have the potential to be particularly useful for tracking the COVID-19 pandemic [51].

Contacts between pathogens and hosts that elicit an immune response can be revealed by the presence of antibodies. Broadly speaking, there are two major classes of antibody, with rising concentrations of immunoglobulin M (IgM) preceding the increase in concentration of immunoglobulin G (IgG). However, for infections by SARS-CoV-2, there is increasing evidence that IgG and IgM appear concurrently [52]. Most serological assays used for understanding viral transmission measure concentrations of IgG. Interpretation of a positive result depends on the availability of detailed knowledge of immune response dynamics and its epidemiological correspondence to the developmental stage of the pathogen, for example the presence of virus shedding in the case of SARS-CoV-2 [53,54]. Serological surveys are common practice in infectious disease epidemiology and have been used to estimate the prevalence of carriers of antibodies, force of infection, and reproduction numbers [55], and in certain circumstances (e.g. for measles) used to infer population immunity to a pathogen [56]. Unfortunately, single serological surveys only provide information about the numbers of individuals who are seropositive at the times the surveys were conducted (as well as information about the individuals tested, such as their age [57]). Although information about temporal changes in infections can be obtained by conducting multiple surveys longitudinally [46,58], the precise timings of infections remain unknown.

Available tests vary in sensitivity and specificity, which can impact the accuracy of predictions made using compartmental models if seropositivity is used to assess the proportion of individuals that are protected from infection or disease. Propagation of uncertainty due to the sensitivity and specificity of the testing procedures and epidemiological interpretation of the immune response are areas that require attention. The possible presence of immunologically silent individuals, as implied by studies of COVID-19 showing that 10–20% of symptomatically infected people have few or no detectable antibodies [59], adds to the known sources of uncertainty. Ambiguities in the interpretation of the biological meaning of testing outcomes and limitations of study designs raise issues related to the identifiability of parameters of interest.

Many compartmental modelling studies have used data on deaths as the main reliable dataset for model fitting. The extent to which seroprevalence data could provide an additional useful input for model calibration, and help in formulating exit strategies, has yet to be ascertained. With the caveats above, one-off or regular assessments of the seroprevalence in the population could be helpful in understanding transmission of SARS-CoV-2 in different populations.

## *1.4 IS GLOBAL ERADICATION OF SARS-COV-2 A REALISTIC POSSIBILITY?*

When $R_0$ is greater than one, an emerging outbreak will either grow to infect a substantial proportion of the population or become extinct before it is able to do so [60–64]. If instead $R_0$ is less than one, the outbreak will almost certainly become extinct before a substantial proportion of the population is infected. If new susceptible individuals are introduced into the population (for example, new susceptible individuals are born), it is possible that that the disease will persist after its first wave and become endemic [65]. These theoretical results can be extended to populations with household and network structure [66,67] and scenarios in which $R_0$ is very close to one [68].

Epidemiological theory and data from multiple different diseases indicate that extinction can be a slow process, often involving a long "tail" of cases with significant random fluctuations (Figure S1). Long epidemic tails can be driven by factors including spatial heterogeneities, such as differences in weather in different countries (potentially allowing an outbreak to persist by surviving in different locations at different times of year) and varying access to treatment in different locations. Regions or countries that eradicate SARS-CoV-2 successfully might experience reimportations from elsewhere [69,70], for example the reimportation of the virus to New Zealand from the UK in June 2020.

At the global scale, smallpox is the only previously endemic human disease to have been eradicated, and extinction took many decades of vaccination. Prevalence and incidence of polio and measles have been reduced substantially through vaccination but both diseases persist. The 2001 Foot and Mouth Disease outbreak in the UK and the 2003 SARS pandemic were new epidemics that were driven extinct without vaccination before they became endemic, but both exhibited long tails before eradication was achieved. The 2014-16 Ebola outbreak in West Africa was eliminated (with vaccination at the end of the outbreak [71]), but eradication took some time with flare ups occurring in different countries [72,73].

Past experience therefore raises the possibility that SARS-CoV-2 may not be driven to complete extinction in the foreseeable future, even if a vaccine is developed and vaccination campaigns can be implemented. As exemplified by the outbreak of Ebola in the Democratic Republic of the Congo that has only recently been declared over [74], there is an additional challenge of assessing whether the virus really is extinct rather than persisting in individuals who do not report disease [72]. There is a distinct possibility that SARS-CoV-2 could become endemic, either persisting in populations with limited access to healthcare or circulating in seasonal outbreaks as the virus evolves. Appropriate communication of these scenarios to the public and policy-makers – particularly the possibility that SARS-CoV-2 may never be eradicated – is of obvious importance.

# 2 HETEROGENEITIES IN TRANSMISSION

## 2.1 HOW MUCH RESOLUTION IS NEEDED WHEN MODELLING HUMAN HETEROGENEITIES?

A common challenge faced by modellers working in outbreak situations is the tension between making models more complex (and possibly therefore seeming more realistic and convincing to stakeholders) and maintaining simplicity (for scientific parsimony when data are sparse and for expediency when predictions are required at very short notice by policy-makers) [75]. How to strike the correct balance is not a settled question, especially given the growing amount of available data on human demography and behaviour. Indeed, outputs of multiple models with different levels of complexity can provide useful and complementary information. Many sources of heterogeneity between individuals (and between populations) exist, including the strong skew of severe COVID-19 outcomes towards the elderly and individuals from specific groups. Here, we focus on two sources of heterogeneity in human populations that must be considered when modelling exit strategies: spatial contact structure and health vulnerabilities.

There has been considerable success in modelling local contact structure, both in terms of spatial heterogeneity (distinguishing local and long-distance contacts) and in local mixing structures such as households and workplaces. However, challenges include tracking pathogen transmission and assessing changes when contact networks are altered. In spatial models with only a small number of near-neighbour contacts, the number of new infections grows slowly, so that each generation of infected individuals is only slightly larger than the previous one. As a result, in those models, $R(t)$ cannot significantly exceed its threshold value of one [76]. In contrast, models accounting for transmission within closely interacting groups explicitly contain a mechanism that has a multiplier effect on the value of $R(t)$ [66]. Another challenge is in modelling the spatiotemporal structure of human populations: the spatial distribution of individuals is important, but long-distance contacts make populations more connected than depicted by simple percolation-type spatial models [76]. Clustering and pair approximation models can capture some aspects of spatial heterogeneities [77], which can result in exponential rather than linear growth in case numbers [78].

While modelling frameworks exist to include almost any kind of spatial stratification, ensuring that model outputs are meaningful for exit strategy planning relies on appropriate calibration with data. This brings in challenges of merging multiple data types with different stratification levels. For example, case notification data may be aggregated at a broad regional level within a country, while mobility data from past surveys might be available at finer scales within regions. Another challenge is to determine the appropriate scale at which to introduce or lift interventions. Although government policies are usually directed at whole populations within relevant administrative units (country-wide or smaller), more effective interventions and exit strategies may exist that target specific parts of the population [79]. Here, modelling can be helpful to account for operational costs and imperfect implementation that will offset expected epidemiological gains.

The structure of host vulnerability to disease is generally reported via risk factors: many factors have been considered, including age, sex and ethnicity [80,81]. From a modelling perspective, a number of open questions exist. To what extent does heterogeneous vulnerability at an individual level affect the impact of exit strategies beyond the reporting of potential outcomes, if at all? Where host vulnerability is an issue, is it necessary to account for considerations other than reported risk factors, as these may be proxies for underlying causes? Another important aspect is that, once communicated to the public, the results of modelling could create behavioural feedback that might help or hinder exit strategies; some sensitivity analyses would be useful. As with the questions around spatial heterogeneity, modelling variations in host vulnerability could improve proposed exit strategies, and modelling might be used to explore how these are targeted and communicated [5]. Finally, heterogeneities in space and vulnerabilities may interact; modelling these may reveal surprises that can be explored further.

## 2.2 WHAT ARE THE ROLES OF NETWORKS AND HOUSEHOLDS IN SARS-COV-2 TRANSMISSION?

In combination with contact tracing, NPIs reduce the opportunity for transmission by breaking up contact networks (closing workplaces, schools, preventing large gatherings), reducing the chance of transmission where links cannot be broken (e.g. wearing masks, sneeze barriers) and identifying infected individuals (temperature checks [82], diagnostic testing [83]). Network models [84,85] aim to split pathogen transmission into opportunity (number of contacts) and transmission probability, using social network data that can be measured directly (through devices such as mobility tracking and contact diaries) and indirectly (through traffic flow and co-occurrence studies). This brings new issues: for example, are observed networks missing key routes of transmission, such as indirect contact via contaminated surfaces, or adding noise in the form of contacts that are low risk [86]? How we measure and interpret contact networks depends on the geographical and social scale of interest (e.g. wider community spread or closed populations such as prisons and care homes; or sub-populations such as workplaces and schools) and the timescale over which we want to use the network to understand or predict infection chains.

In reality, individuals belong to households, children attend schools and adults mix in workplaces as well as in social contexts. This has led to the development of household models [66,87–90], multilayer networks [91], bipartite networks [92,93] and networks that are geographically- and socially-embedded to reflect location and travel habits [94]. These modelling tools can play a key role in understanding and monitoring transmission, and exploring what-if scenarios, at the point of exiting a lockdown: in particular, they can inform whether or not, and how quickly, households or local networks merge to form larger and possibly denser contact networks in which local outbreaks can emerge. Variations between regions and socio-economic factors can also be explored.

Contact tracing, followed by isolation or treatment of infected contacts, is a well-established method of disease control. The structure of the contact network is important in determining whether or not contact tracing will be successful. For example, contact tracing in clustered networks is known to be more effective than in equivalent non-clustered networks [95,96], since a potentially infected contact can be traced from multiple different sources. Knowledge of the contact network enhances understanding of the correlation structure that emerges as a result of the epidemic. The first wave of an epidemic will typically infect many of the highly connected nodes and will move slowly to less connected parts of the network leaving behind islands of removed and susceptible individuals. This can lead to a correlated structure of susceptible and recovered nodes that may make the networks less vulnerable to later epidemic waves [97], and has implications for herd immunity (Section 1.2).

In heterogeneous populations, relatively few very well-connected people can have a large impact on the spread of a pathogen and be major hubs for transmission. Such individuals are often referred to as super-spreaders [98,99] and some theoretical approaches to controlling epidemics are based on targeting them [100]. However, particularly for respiratory diseases, whether specific individuals can be classified as potential super-spreaders, or instead whether any infected individual has the potential to generate super-spreading events, is debated [37,101,102].

As control policies are gradually lifted, the disrupted contact network will start to form again. Understanding how proxies for social networks (which can be measured in near-real time using mobility data, electronic sensors or trackers) relate to transmission requires careful consideration. Using observed contacts to predict virus spread might be successful if these quantities are heavily correlated, but one aim of NPIs should be at least a partial decoupling of the two, so that society can reopen but transmission remains controlled. The extent to which this is possible is unclear and is likely to vary between regions. Currently, a key empirical and theoretical challenge is to understand how households are connected and how this is affected by school opening (Section 2.3). An important area for further research is to improve our understanding of the role of within-household transmission in the ongoing COVID-19 pandemic. In particular, do sustained infection chains within households lead to amplification of infection rates between households despite lockdowns aimed at minimising between-household transmission?

Even for relatively well-studied household models, theoretical development of methods accommodating time-varying parameters such as variable adherence to household-based policies and/or compensatory behaviour would be valuable to inform future control policies. It would be useful to compare interventions and de-escalation procedures in different countries to gain insight into: how contact and transmission networks vary between regions; the role of different household structures in the spread and severity of outcomes (accounting for different household sizes and age-structures); the cost-effectiveness of different policies, such as household-based isolation and quarantine in the UK compared to out-of-household quarantine in Australia and Hong Kong. First Few X (FFX) studies [103,104], now adopted in several countries, provide the opportunity not only to improve our understanding of critical epidemiological characteristics (such as incubation periods, generation intervals and the roles of asymptomatic and presymptomatic transmission) but also to make many of the comparisons just outlined.

## 2.3 WHAT IS THE ROLE OF CHILDREN IN SARS-COV-2 TRANSMISSION?

An early intervention implemented in many countries was school closure, which is frequently used during influenza pandemics [105,106]. Further, playgrounds were closed and strict social distancing has kept children separated. However, the role of children in SARS-CoV-2 transmission is unclear. Early signs from Wuhan (China), echoed elsewhere, showed many fewer cases in under 20s than might have been expected. There are three aspects of the role of children in transmission: i) susceptibility to infection; ii) infectiousness once infected; iii) propensity to develop disease if infected [107–109]. Evidence for variation in age-specific susceptibility to infection and infectiousness is mixed, with infectiousness the more difficult to quantify. However, evidence is emerging of lower susceptibility to infection in children compared to adults [110], although the mechanism underlying this is unknown and it may not be generalisable to all settings. Once infected, children appear to have a milder course of infection than adults, and it has been suggested that children have a higher probability of a fully subclinical course of infection.

Reopening schools is of clear importance both in ensuring equal access to education and enabling caregivers to return to work. However, the risk of transmission within schools and the potential impact on the rate of community transmission needs to be understood so that policy-makers can balance the potential benefits and harms. As schools begin to reopen there are key questions that models can help with, and major knowledge gaps that prevent clear answers. The most pressing question at a regional and national level is the extent to which school restarting will affect population-level transmission, characterised by $R(t)$ (Section 1.1). Clearer quantification of the role of children could have come from analysing the effects of school closures in different countries in February and March, but as described in the Introduction, closures generally coincided with many other interventions and so it has proved difficult to unpick the effects of individual measures [7]. Almost all schools in Sweden stayed open to under-16s (with the exception of one school that closed for two weeks [111]), and schools in some other countries are beginning to reopen with social distancing measures in place, providing a potential opportunity to improve our understanding of within-school transmission. Models can also inform the design of studies to generate the data required to answer key questions.

The effect of opening schools on $R(t)$ also depends on other changes in the rest of the community. Children, teachers, and support staff are members of households, and lifting restrictions may therefore affect all members. Modelling school reopening must account for all changes in contacts of household members [112], noting that the impact on $R(t)$ may depend on the other interventions in place at that time. The relative risk of restarting different school years (or even universities) does not affect the population $R(t)$ straightforwardly, since older children tend to live with adults who are older (compared to younger children), and households with older individuals are at greater risk of seeing severe outcomes compared to households with younger ones. Thus, decisions about which age groups return to school first and how they are grouped at school must balance the risks of transmission between children at school, transmission to and between their teachers, and transmission to and within the households of those children.

Return to school affects the number of physical contacts that teachers and support staff have. Schools will not be the same environments as prior to lockdown, since physical distancing measures will be in place. These include smaller classes and changes in layout, plus increased hygiene measures to decrease transmission. This is critical to reduce SARS-CoV-2 spread between teachers as well as from teachers to children (and, perhaps to a lesser extent, from children to teachers). Some teachers may be unlikely to return to school because of the presence of underlying conditions and a need to "shield", and if there is transmission within schools, there may be absenteeism following infection. Models must therefore consider the different effects on transmission of pre- and post-lockdown school environments. Post-lockdown, with strong social distancing in place in the wider community, reopening schools could link subcommunities of the population together, and models can be used to estimate the wider effects on population transmission as well as those within schools themselves. These estimates are likely to play a central role in decisions surrounding when and how to reopen schools in different countries.

## 2.4 THE PANDEMIC IS SOCIAL: HOW CAN WE MODEL THAT?

As the pandemic progresses, so does the need for different modelling approaches that account for population structure and heterogeneity. While these effects can be approximated in standard compartmental epidemiological models [2,72,113], such models can become highly complex and cumbersome to specify and solve as more sources of heterogeneity are introduced. An alternative modelling paradigm is agent-based modelling. Agent-based models (ABM) allow complex systems such as societies to be represented, using virtual agents programmed to have behavioural and individual characteristics (age, sex, ethnicity, income, employment status, etc.) as well as the capacity to interact with other agents [114]. In addition, ABM can include the effects of societal-level factors such as the influence of social media, regulations and laws, and community norms. In more sophisticated ABM, agents can anticipate and react to scenarios, and learn by trial and error or by imitation. ABM represent a way to model systems in which there are feedbacks, tipping points, the emergence of higher-level properties from the actions of individual agents, adaptation, and multiple scales of organisation – all features of the COVID-19 pandemic and societal reactions to it.

While ABM arise from a different tradition, they can incorporate the insights of compartmental models; for example, agents must transition through disease states (or compartments) such that the mean transition rates correspond to the rates that quantify flows in compartmental models. However, building an ABM that represents a population on a national scale is a huge challenge and one that is unlikely be accomplished in a timescale useful for the current pandemic. ABM often include many parameters, leading to challenges of model parameterisation and a requirement for careful uncertainty quantification and sensitivity analyses to different inputs. On the other hand, useful ABM do not have to be all-encompassing. There are already several models accessible to the public that illustrate the effects of policies such as social distancing on small simulated populations. These models can be very helpful as "thought experiments" to identify the potential effects of candidate policies such as school re-opening, the consequences of non-compliance with government edicts and the impacts of restrictions on long-distance travel, amongst others.

There are two areas where action should be taken, both of which are long-term and intended to assist in dealing with the almost inevitable next pandemic rather than this one. First, more data about people's ordinary behaviour are required: what individuals do each day (through time-use diaries), whom they meet (possibly through mobile phone data, assuming consent can be obtained), and how they understand and act on government regulation, social media influences and broadcast information [115]. Second, a large, modular ABM should be built that represents heterogeneities in populations and that is properly calibrated as a social "digital twin" of our own society, with which we can carry out virtual policy experiments. Had these developments occurred before, they would have been useful in the current situation. As a result, if these are done now, they will aid the planning of exit strategies in future.

# 3  DATA NEEDS AND COMMUNICATING UNCERTAINTY

## 3.1  *WHAT ARE THE ADDITIONAL CHALLENGES OF DATA LIMITED SETTINGS?*

In most countries, criteria for ending COVID-19 lockdowns rely on tracking trends in numbers of confirmed cases and deaths, and assessments of transmissibility (Section 1.1). In this section, we focus specifically on issues in determining when interventions should be relaxed in LMICs, although we note that many of these issues apply everywhere. Perhaps surprisingly, many concerns relating to data availability and reliability (e.g. lack of clarity about sampling frames) remain in countries worldwide. Other difficulties have also been experienced in many countries throughout the pandemic (e.g. shortages of vital supplies, perhaps due in developed countries to previous emphasis on healthcare system efficiency rather than pandemic preparedness [116]).

In many countries, data about the COVID-19 pandemic and about the general population and context can be unreliable or lacking. Because of limited healthcare access and utilisation, there can be fewer opportunities for diagnosis and subsequent confirmation of cases in LMICs compared to other settings, unless there are active programmes [117]. Distrust can make monitoring programmes difficult, and complicate control activities like test-trace-isolate campaigns [118,119]. Other options for monitoring – such as assessing excess disease from general reporting of acute respiratory infections or influenza-like-illness – require historical baselines that may not exist [120,121]. In general, while many LMICs will have a well-served fraction of the population, dense peri-urban and informal settlements are typically outside that population and may rapidly become a primary concern for transmission [122]. Since confirmed case numbers in these populations are unlikely to provide an accurate representation of the underlying epidemic, reliance on alternative measures such as clinically diagnosed cases may be necessary to understand the epidemic trajectory. Some tools for rapid assessment of mortality in countries where the numbers of COVID-19 related deaths are hard to track are starting to become available [123].

In addition to the challenges in understanding the pandemic in these settings, metrics on health system capacity (including resources such as beds and ventilators), as needed to set targets for control, are often poorly documented [124]. Furthermore, the economic hardships and competing health priorities in low-resource settings change the objectives of lifting restrictions – for example, hunger due to loss of jobs and changes in access to routine health care (e.g. HIV services and childhood vaccinations) as a result of lockdown have the potential to cost many lives in themselves, both in the short- and long-term [125,126], and this must be accounted for when deciding how to relax COVID-19 public health measures. To assess the costs and benefits of lifting restrictions appropriately, additional data on these conditions may be required. In many settings, these data are currently unavailable.

These challenges suggest three key elements of modelling efforts for moving forward in data-limited settings: i) identification of policy responses that are robust to missing information; ii) value-of-information analyses to prioritise additional data collection and curation efforts; iii) development of methods that use metadata to interpret epidemiological patterns.

In settings where additional data collection is not affordable, models may provide a clearer picture by incorporating available metadata, such as testing and reporting rates through time, sample backlogs, and suspected COVID-19 cases based on syndromic surveillance. By identifying the most informative data, modelling could encourage countries to share available data more widely. For example, in high incidence settings, burial reports and death certificates may be available, and these data can provide information on the demographics that influence the infection fatality rate. These can in turn reveal potential COVID-19 deaths classified as other causes and hence missing from COVID-19 attributed death notifications.

In general, supporting LMICs calls for creativity in the data that might be incorporated in models and in the response activities that are undertaken. Some LMICs have managed the COVID-19 pandemic successfully so far (e.g. Vietnam, as well as Trinidad and Tobago [127]). However, additional support in LMICs is required: data limited settings represent uniquely high stakes.

## 3.2 WHICH DATA SHOULD BE COLLECTED AS COUNTRIES EMERGE FROM LOCKDOWN, AND WHY?

Identifying the effects of the different components of lockdown is important to understand how – and in which order – interventions should be released. The impact of previous measures must be understood both to inform policy in real-time and to ensure that lessons can be learnt from the current pandemic.

Models vary from those that include few parameters but can offer powerful and robust insights into the potential impacts of different strategies, to highly complex simulations aiming to capture all nuances affecting transmission. Complex simulations are often sensitive to uncertainties in the many assumed parameters and model structure. Ultimately, all models require information to make their predictions relevant to the ongoing pandemic. Data from PCR tests for presence of active virus and serological tests for antibodies, together with data on COVID-19 related deaths, are freely available via a number of internet sites (e.g. [128]). However, metadata associated with testing protocols (e.g. reason for testing, type of test, breakdowns by age and underlying health conditions) and the definition of COVID-19 related death, which are needed to quantify sources of potential bias and parameterise models correctly, are often unavailable. Data from individuals likely to have been exposed to the virus (e.g. within households of known infected individuals), but who may or may not have contracted it themselves, are also useful for model parameterisation [129]. New sources of data exist, ranging from tracking data from mobile phones [130] to social media surveys [131] and details of interactions with public health providers [132]. Although potentially valuable, these data sources bring with them biases that are not always understood perfectly. These types of data are also often subject to data protection and/or costly fees, meaning that they are not readily available to many modelling groups. Mixing patterns by age were reasonably well-characterised before the current pandemic [133,134] (particularly for adults of different ages) and have been used extensively in existing models. However, there are gaps in these data and uncertainty in the impacts that different interventions have had on mixing. Predictive models for policy tend to make broad assumptions about the effects of elements of social distancing [135], although results of studies that attempt to estimate effects in a more data-driven way are beginning to emerge [136]. The future success of modelling efforts to understand when controls should be relaxed or tightened depends critically on whether, and how accurately as well as how quickly, the effects of different elements of lockdown can be parameterised.

Given the many differences in lockdown implementation between countries, cross-country comparisons offer an opportunity to estimate the effects on transmission of each component of lockdown [7]. However, there are many challenges in comparing SARS-CoV-2 dynamics in different countries. Alongside variability in the timing, type and impact of interventions, the numbers of importations from elsewhere will vary between countries [69,137]. Underlying differences in mixing, behavioural changes in response to the pandemic, household structures, occupations and distributions of ages and co-morbidities are likely to be important but uncertain drivers of transmission patterns. A current research target is to understand the role of weather and climate in SARS-CoV-2 transmission and severity [138]. Many analyses across and within countries highlight potential correlations between environmental variables and transmission [139–144], although sometimes by applying ecological niche modelling frameworks that are perhaps ill-suited for modelling a rapidly spreading pathogen [145–147]. Assessments of the interactions between weather and viral transmissibility are facilitated by the availability of extensive datasets describing weather patterns, such as the European Centre for Medium-Range Weather Forecasts ERA5 dataset [148] and simulations of the Community Earth System Model that can be used to estimate the past, present and future values of meteorological variables worldwide [149]. It is likely that temperature, humidity and precipitation affect the survival of SARS-CoV-2 outside the body, and prevailing weather conditions could, in theory, tip $R(t)$ above or below one. However, the effects of these factors on transmission have not been established conclusively, and the impact of seasonality on short- or long-term SARS-CoV-2 dynamics is likely to depend on other factors including the timing and impact of interventions, and the dynamics of immunity [46,150]. At present, it is hard to separate the effects of the weather on virus survival from other factors including behavioural changes in different seasons [151]. The challenge of disentangling the impact of variations in weather on transmission from other epidemiological drivers in different locations is therefore a complex open problem.

In seeking to understand and compare COVID-19 data from different countries, there is a need to coordinate the design of epidemiological studies, involving longitudinal data collection and case-control studies. This will help enable models to track the progress of the epidemic and the impacts of control policies internationally. It will also allow more refined conclusions than those that follow from population data alone. Countries with substantial epidemiological modelling expertise should support epidemiologists elsewhere with standardised protocols for collecting data and using models to inform policy. There is also a need to share models to be used "in the field" in different settings. Collectively, these efforts will ensure that models are parameterised as realistically as possible for the particular settings in which they are to be used. In turn, as interventions are relaxed, this will allow us to detect the earliest possible reliable signatures of a resurgence in cases should it occur, leading to an unambiguous characterisation of when it is necessary for some interventions to be reintroduced.

## 3.3   HOW SHOULD MODEL AND PARAMETER UNCERTAINTY BE COMMUNICATED?

SARS-CoV-2 transmission models have played a crucial role in shaping policies in different countries, and their predictions (and the scenarios that they have been used to explore) have been a regular feature of media coverage of the pandemic [135,152]. Understandably, both policy-makers and journalists generally prefer single "best guess" figures from models, rather than a range of plausible values. However, the ranges of outputs that modellers provide include important information about the variety of possible scenarios and guard against over-interpretation of model results. Not presenting information about uncertainty can convey a false confidence in predictions, and it is critical that modellers present uncertainty in a way that is understandable and useful for policy-makers and the public [75].

There are numerous and often inextricable ways in which uncertainty enters the modelling process. Any model includes assumptions that inevitably vary according to judgements regarding which features should be included in the model [1,94] and which datasets are used to inform the model [153]. Within any model, ranges of parameter values can be considered to allow for uncertainty about clinical characteristics of COVID-19 (e.g. the infectious period and case fatality rate) [154]. Alternative initial conditions (e.g. numbers and locations of imported cases seeding national outbreaks, or levels of population susceptibility) can be considered. In modelling exit strategies, when surges in cases starting from small numbers may occur and where predictions will depend on characterising epidemiological parameters as accurately as possible, stochastic models may be of particular importance. Not all the uncertainty arising from such stochasticity will be reduced by collecting more data; it is inherent to the process.

Where models have been developed for similar purposes, formal methods of comparison can be applied, but in epidemiological modelling, models often have been developed to address different questions, possibly involving "what-if?" scenarios, in which case only qualitative comparisons can be made. The ideal outcome for policy-making is when different models generate similar conclusions, demonstrating robustness to the detailed assumptions involved. Where there is a narrowly defined requirement, such as short-term predictions of cases and deaths, more tractable tools for comparing the outputs from different models in real-time would be valuable. One possible approach is to assess and compare the models' past performance at making predictions [33,155]. The use of ensemble estimates, most commonly applied for forecasting disease trajectories, is a way to synthesise multiple models' predictions into a single estimate [156,157]. The assessment and comparison of past performance can then be used to weight models in the ensemble. Such approaches typically lead to improved point and variance estimates.

To deal with parameter uncertainty, a common approach is to perform sensitivity analyses in which model parameters are repeatedly sampled from a range of plausible values, and the resulting model predictions compared; both classical and Bayesian statistical approaches can be employed [158–160]. Methods of uncertainty quantification provide a framework in which uncertainties with regard to model structure, values of epidemiological parameters, and data can be considered together. In practice, there is usually only a limited number of possible policies that can be implemented. An important question is often whether or not the optimal policy can be identified given the uncertainties we have described, and decision analyses can be helpful for this purpose [161,162].

In summary, communication of uncertainty to policy-makers and the general public remains a challenging area. Different levels of detail may be required for different audiences. There are many subtleties: for instance, almost any epidemic model can provide an acceptable fit to data in the early phase of an outbreak, since models almost invariably predict exponential growth. This can induce an artificial belief that the model must be based on sensible underlying assumptions, and the true uncertainty about such assumptions has vanished. Clear presentation of data is critical. For example, it is important not simply to present data on the numbers of cases, but also include information about the numbers of individuals who have been tested. In addition, clear statements of the individual values used to calculate quantities such as the case fatality rate are vital, so that studies can be interpreted and compared correctly [163,164]. Going forwards, improved communication of model and parameter uncertainty is essential as models are used to predict the effects of different exit strategies.

## SUMMARY AND DISCUSSION

In this article, we have highlighted a number of ongoing challenges in modelling the COVID-19 pandemic, and uncertainties faced by most countries devising lockdown exit strategies. It is important, however, to put these issues into context: at the start of 2020 the virus was unknown, and its pandemic potential only became apparent at the end of January. The speed with which the scientific and public health communities came together to tackle this challenge and the openness in sharing data, methods and analyses are unprecedented. At very short notice, epidemic modellers were able to mobilise a substantial workforce – mostly on a voluntary basis – and state-of-the-art computational models. Far from the rough-and-ready tools sometimes depicted in the media, the modelling effort deployed since January is a collective and multi-pronged effort benefitting from years of experience of outbreak modelling, often combined with long-term engagement with public health agencies and policy-makers.

Drawing on this collective expertise, the virtual meeting convened in mid-May by the Isaac Newton Institute generated a clear overview of the steps needed to improve and validate the scientific advice to guide lockdown exit strategies. Importantly, the roadmap outlined in this paper is meant to be feasible within the lifetime of the pandemic. Unlike some scientific fields, infectious disease epidemiology does not have the luxury of waiting for all data to become available before fully validated models must be developed. As discussed here, the solution lies in using diverse and flexible modelling frameworks that can be revised and improved iteratively as more data become available. Equally important is the ability to assess the data critically and bring together evidence from multiple fields: numbers of cases and deaths reported by regional or national authorities only represent a single source of data, and expert knowledge is even required to interpret these data correctly.

In this spirit, our first recommendation is to improve estimates of key epidemiological parameters. This requires close collaboration between epidemic modellers and the individuals and organisations that collect epidemic data, so that the caveats and assumptions on each side are clearly presented and understood. That is a key message from the first section of this study, in which the relevance of theoretical concepts and model parameters in the real world was demonstrated: far from ignoring the complexity of the pandemic, models draw from different sources of expertise to make sense of imperfect observations. By acknowledging the simplifying assumptions of models, we can assess their relative impacts and validate or replace them as new evidence becomes available.

Our second recommendation is to seek to understand important sources of heterogeneity that appear to be driving the pandemic and its response to interventions. Agent-based modelling represents one possible framework for modelling complex dynamics, but standard epidemic models can also be extended to include age groups or any other relevant strata in the population as well as spatial structure. Network models provide computationally efficient approaches to capture different types of epidemiological and social interactions. Importantly, many modelling frameworks provide avenues for collaboration with other fields, such as the social sciences.

Our third and final recommendation regards the need to focus on data requirements, particularly (although not exclusively) in resource limited settings such as LMICs. Understanding the data required for accurate predictions in different countries requires close communication between modellers and governments, public health authorities and the general public. While this pandemic casts a light on social inequalities between and within countries, modellers have a crucial role to play in sharing knowledge and expertise with those who need it most. In LMICs, cost-effective guidance can be provided by models validated with global data. During the pandemic so far, countries that might be considered similar in many respects have often differed in their policies; either in the choice or the timing of restrictions imposed on their respective populations. Models are important for drawing reliable inferences from global comparisons of the relative impacts of different control measures. All too often, national death tolls have been used for political purposes in the media, attributing the apparent success or failure of particular countries to specific policies without presenting any convincing evidence. Modellers must work closely with policy-makers, journalists and social scientists to improve the communication of rapidly changing scientific knowledge while conveying the multiple sources of uncertainty in a meaningful way.

We are now moving into a stage of the COVID-19 pandemic in which data collection and novel research to inform the modelling issues discussed here are both possible and essential for global health. These are international challenges that require an international collaborative response from diverse scientific communities, which we hope that this article will stimulate. This is of critical importance, not only to tackle this pandemic but also to improve the response to future outbreaks of emerging infectious diseases.

## References

1. Prem K, Liu Y, Russell TW, Kucharski AJ, Eggo RM, Davies N, et al. The effect of control strategies to reduce social mixing on outcomes of the COVID-19 epidemic in Wuhan, China: a modelling study. Lancet Public Heal. 2020;5: e261–e270.
2. Thompson RN. Epidemiological models are important tools for guiding COVID-19 interventions. BMC Med. 2020;18: 152.
3. Leung K, Wu JT, Liu D, Leung GM. First-wave COVID-19 transmissibility and severity in China outside Hubei after control measures, and second-wave scenario planning: a modelling impact assessment. Lancet. 2020;395: 1382–1393.
4. Rawson T, Brewer T, Veltcheva D, Huntingford C, Bonsall MB. How and when to end the COVID-19 lockdown: an optimisation approach. Front Public Heal. 2020;8: 262.
5. van Bunnik BAD, Morgan ALX, Bessell P, Calder-Gerver G, Zhang F, Haynes S, et al. Segmentation and shielding of the most vulnerable members of the population as elements of an exit strategy from COVID-19 lockdown. medRxiv. 2020.
6. White LF, Moser CB, Thompson RN, Pagano M. Statistical estimation of the reproductive number from case notification data: a review. Am J Epidemiol. 2020.
7. Flaxman S, Mishra S, Gandy A, Unwin HJT, Coupland H, Mellan TA, et al. Report 13: Estimating the number of infections and the impact of non-pharmaceutical interventions on COVID-19 in 11 European countries. 2020. Available: www.imperial.ac.uk/media/imperial-college/medicine/mrc-gida/2020-03-30-COVID19-Report-13.pdf
8. Abbott S, Hellewell J, Thompson RN, Sherratt K, Gibbs HP, Bosse NI, et al. Estimating the time-varying reproduction number of SARS-CoV-2 using national and subnational case counts. Wellcome Open Res. 2020;5: 112.

9. Cowling BJ, Ali ST, Ng TW, Tsang TK, Li JC, Fong MW, et al. Impact assessment of non-pharmaceutical interventions against coronavirus disease 2019 and influenza in Hong Kong: an observational study. Lancet Public Heal. 2020;5: e279–e288.

10. Gostic KM, McGough L, Baskerville E, Abbott S, Joshi K, Tedijanto C, et al. Practical considerations for measuring the effective reproductive number, Rt. medRxiv. 2020.

11. Haydon DT, Chase-Topping M, Shaw DJ, Matthews L, Friar JK, Wilesmith J, et al. The construction and analysis of epidemic trees with reference to the 2001 UK foot-and-mouth outbreak. Proc R Soc B Biol Sci. 2003;270: 121–127.

12. Birrell P, Blake J, van Leeuwen E, De Angelis D, Joint PHE Modelling Cell, MRC Biostatistics Unit COVID-19 Working Group. Nowcasting and Forecasting Report. 2020 [cited 21 May 2020]. Available: https://www.mrc-bsu.cam.ac.uk/now-casting/

13. Government Office for Science. Government publishes latest R number. 2020. Available: https://www.gov.uk/government/news/government-publishes-latest-r-number

14. UK Government Cabinet Office. Our plan to rebuild: The UK government's COVID-19 recovery strategy. 2020.

15. Cauchemez S, Boëlle P, Thomas G, Valleron AJ. Estimating in real time the efficacy of measures to control emerging communicable diseases. 2006. 164: 591–597.

16. Wallinga J, Teunis P. Different epidemic curves for severe acute respiratory syndrome reveal similar impacts of control measures. Am J Epidemiol. 2004;160: 509–516.

17. Cori A, Ferguson NM, Fraser C, Cauchemez S. A new framework and software to estimate time-varying reproduction numbers during epidemics. Am J Epidemiol. 2013;178: 1505–12.

18. Thompson RN, Stockwin JE, Gaalen RD Van, Polonsky JA, Kamvar ZN, Demarsh PA, et al. Improved inference of time-varying reproduction numbers during infectious disease outbreaks. Epidemics. 2019;19: 100356.

19. Lemaitre JC, Perez-Saez J, Azman AS, Rinaldo A, Fellay J. Assessing the impact of non-pharmaceutical interventions on SARS-CoV-2 transmission in Switzerland. Swiss Med Wkly. 2020;150: w20295.

20. Nishiura H, Chowell G. The effective reproduction number as a prelude to statistical estimation of

time-dependent epidemic trends. Mathematical and Statistical Estimation Approaches in Epidemiology. 2009. pp. 103–121.

21. Stadler R, Kuhnert D, Bonhoeffer S, Drummond AJ. Birth–death skyline plot reveals temporal changes of epidemic spread in HIV and hepatitis C virus (HCV). PNAS. 2013;110: 228–233.

22. Volz EM, Didelot X. Modeling the growth and decline of pathogen effective population size provides insight into epidemic dynamics and drivers of antimicrobial resistance. Syst Biol. 2018;67: 719–728.

23. Pybus OG, Charleston MA, Gupta S, Rambaut A, Holmes EC, Harvey PH. The epidemic behavior of the hepatitis C virus. Science (80- ). 2001;22: 2323–2325.

24. Parag K V, Donnelly CA. Adaptive estimation for epidemic renewal and phylogenetic skyline models. Syst Biol. 2020;1: syaa035.

25. Donker T, van Boven M, van Ballegooijen WM, van't Klooster TM, Wielders CC, Wallinga J. Nowcasting pandemic influenza A/H1N1 2009 hospitalizations in the Netherlands. Eur J Epidemiol. 2011;26: 195–201.

26. Flaxman S, Mishra S, Gandy A, Unwin HJT, Mellan TA, Coupland H, et al. Estimating the effects of non-pharmaceutical interventions on COVID-19 in Europe. Nature. 2020.

27. Richardson WH. Bayesian-based iterative method of image restoration. J Opt Soc Am. 1972;62: 55–59.

28. Lucy LB. An iterative technique for the rectification of observed distributions. Astron J. 1974;79: 745–754.

29. Goldstein E, Dushoff J, Ma J, Plotkin JB, Earn DJD, Lipsitch M. Reconstructing influenza incidence by deconvolution of daily mortality time series. PNAS. 2009;106: 21829.

30. Seaman S, De Angelis D, MRC-BSU COVID-19 Working Group, PHE Modelling Cell. Adjusting COVID-19 deaths to account for reporting delay. 2020. Available: https://www.mrc-bsu.cam.ac.uk/wp-content/uploads/2020/05/Adjusting-COVID-19-deaths-to-account-for-reporting-delay.pdf

31. Vollmer MAC, Mishra S, Unwin HJT, Gandy A, Mellan TA, Bradley V, et al. Report 20: Using mobility to estimate the transmission intensity of COVID-19 in Italy: A subnational analysis with future scenarios. 2020. Available: www.imperial.ac.uk/media/imperial-college/medicine/mrc-gida/2020-05-04-COVID19-Report-20.pdf

32. Svensson A. A note on generation times in epidemic models. Math Biosci. 2007;208: 300–311.

33. Parag K V, Donnelly C. Using information theory to optimise epidemic models for real-time prediction and estimation. PLoS Comput Biol. 2020;16: e1007990.

34. Parag K V, Donnelly CA, Jha R, Thompson RN. An exact method for quantifying the reliability of end-of-epidemic declarations in real time. medRxiv. 2020.

35. Golding N, Shearer FM, Moss R, Dawson P, Gibbs L, Alisic E, et al. Estimating temporal variation in transmission of COVID-19 and adherence to social distancing measures in Australia. 2020. Available: https://www.doherty.edu.au/uploads/content_doc/Technical_report_15_Maypdf.pdf

36. Glass K, Mercer GN, Nishiura H, McBryde ES, Becker NG. Estimating reproduction numbers for adults and children from case data. J R Soc Interface. 2011;8: 1248–1259.

37. Lloyd-Smith JO, Schreiber SJ, Kopp PE, Getz WM. Superspreading and the effect of individual variation on disease emergence. Nature. 2005;438: 355–359.

38. Woolhouse MEJ, Dye C, Etard JF, Smith T, Charlwood JD, Garnett GP, et al. Heterogeneities in the transmission of infectious agents: Implications for the design of control programs. PNAS. 1997;94: 338–342.

39. Endo A, CMMID COVID-19 working group, Abbott S, Kucharski AJ, Funk S. Estimating the overdispersion in COVID-19 transmission using outbreak sizes outside China. Wellcome Open Res. 2020;5: 67.

40. Ferretti L, Wymant C, Kendall M, Zhao L, Nurtay A, Abeler-Dorner L, et al. Quantifying SARS-CoV-2 transmission suggests epidemic control with digital contact tracing. Science (80- ). 2020; eabb6936.

41. Thompson RN, Lovell-Read FA, Obolski U. Time from symptom onset to hospitalisation of coronavirus disease 2019 (COVID-19) cases: Implications for the proportion of transmissions from infectors with few symptoms. J Clin Med. 2020;9: 1297.

42. Du Z, Xu X, Wu Y, Wang L, Cowling BJ, Meyers LA. Serial interval of COVID-19 among publicly reported confirmed cases. Emerg Infect Dis. 2020;26: 1341–1343.

43. Ganyani T, Kremer C, Chen D, Torneri A, Faes C, Wallinga J, et al. Estimating the generation interval for coronavirus disease (COVID-19) based on symptom onset data, March 2020. Eurosurveillance. 2020;25: 2000257.

44. Zhao S. Estimating the time interval between transmission generations when negative values occur in the serial interval data: using COVID-19 as an example. Math Biosci Eng. 2020;17: 3512.

45. Pellis L, Scarabel F, Stage HB, Overton CE, Chappell LH, Lythgoe KA, et al. Challenges in control of Covid-19: short doubling time and long delay to effect of interventions. arXiv. 2020.

46. Kissler SM, Tedijanto C, Goldstein E, Grad YH, Lipsitch M. Projecting the transmission dynamics of SARS-CoV-2 through the postpandemic period. Science (80- ). 2020;368: 860–868.

47. Britton T, Ball F, Trapman P. A mathematical model reveals the influence of population heterogeneity on herd immunity to SARS-CoV-2. Science (80- ). 2020.

48. Gomes MGM, Aguas R, Corder RM, King JG, Langwig KE, Souto-Maior C, et al. Individual variation in susceptibility or exposure to SARS-CoV-2 lowers the herd immunity threshold. medRxiv. 2020.

49. UK Government Office for National Statistics. Coronavirus (COVID-19) Infection Survey pilot: England, 21 May 2020. 2020. Available: https://www.ons.gov.uk/peoplepopulationandcommunity/healthandsocialcare/conditionsanddiseases/bulletins/coronaviruscovid19infectionsurveypilot/england21may2020

50. Metcalf CJE, Farrar J, Cutts FT, Basta NE, Graham AL, Lessler J, et al. Use of serological surveys to generate key insights into the changing global landscape of infectious disease. Lancet. 2016;388: 728–730.

51. Eckerle I, Meyer B. SARS-CoV-2 seroprevalence in COVID-19 hotspots. Lancet. 2020.

52. Borremans B, Gamble A, Prager KC, Helman SK, McClain AM, Cox C, et al. Quantifying antibody kinetics and RNA shedding during early-phase SARS-CoV-2 infection. medRxiv. 2020.

53. Altmann DM, Douek DC, Boyton RJ. What policy makers need to know about COVID-19 protective immunity. Lancet. 2020;395: 1527–1529.

54. Sethuraman N, Jeremiah SS, Ryo A. Interpreting diagnostic tests for SARS-CoV-2. JAMA. 2020.

55. Hens N, Aerts M, Faes C, Shkedy Z, Lejeune O, Van Damme P, et al. Seventy-five years of

estimating the force of infection from current status data. Epidemiol Infect. 2010;138: 802–812.

56. Winter AK, Martinez ME, Cutts FT, Moss WJ, Ferrari MJ, Mckee A, et al. Benefits and challenges in using seroprevalence data to inform models for measles and rubella elimination. J Infect Dis. 2018;218: 355–364.

57. Keiding N. Age-Specific Incidence and Prevalence: A Statistical Perspective. J R Stat Soc Ser A. 1991;154: 371–412.

58. Pollán M, Pérez-Gómez B, Pastor-Barriuso R, Oteo J, Hernán MA, Pérez-Olmeda M, et al. Prevalence of SARS-CoV-2 in Spain (ENE-COVID): a nationwide, population-based seroepidemiological study. Lancet. 2020;6736: 31483–31485.

59. Tan W, Lu Y, Zhang J, Wang J, Dan Y, Tan Z, et al. Viral kinetics and antibody responses in patients with COVID-19. medRxiv. 2020.

60. Thompson RN, Gilligan CA, Cunniffe NJ. When does a minor outbreak become a major epidemic? Linking the risk from invading pathogens to practical definitions of a major epidemic. bioRxiv. 2019.

61. Andersson H, Britton T. Stochastic epidemic models and their statistical analysis. Springer; 2000.

62. Thompson RN. Novel coronavirus outbreak in Wuhan, China, 2020: Intense surveillance is vital for preventing sustained transmission in new locations. J Clin Med. 2020;9: 498.

63. Craft ME, Beyer HL, Haydon DT. Estimating the probability of a major outbreak from the timing of early cases: an indeterminate problem? PLoS One. 2013;8: e57878.

64. Thompson RN, Gilligan CA, Cunniffe NJ. Detecting presymptomatic infection is necessary to forecast major epidemics in the earliest stages of infectious disease outbreaks. PLoS Comput Biol. 2016;12: e1004836.

65. Ballard P, Bean N, Ross J. Intervention to maximise the probability of epidemic fade-out. Math Biosci. 2017;293: 1–10.

66. Ball F, Mollison D, Scalia-Tomba G. Epidemics with two levels of mixing. Ann Appl Probab. 1997;7: 46–89.

67. Bailey NTJ. The use of chain-binomials with a variable chance of infection for the analysis of intra-household epidemics. Biometrika. 1953;40: 279–286.

68. Brightwell G, House T, Luczak M. Extinction times in the subcritical stochastic SIS logistic epidemic. J Math Biol. 2018;77: 455–493.

69. Daon Y, Thompson RN, Obolski U. Estimating COVID-19 outbreak risk through air travel. J Travel Med. 2020; taaa093.

70. Linka K, Rahman P, Goriely A, Kuhl E. Is it safe to lift COVID-19 travel bans? The Newfoundland story. medRxiv. 2020.

71. Henao-Restrepo AM, Longini IM, Egger M, Dean NE, Edmunds WJ, Camacho A, et al. Efficacy and effectiveness of an rVSV-vectored vaccine expressing Ebola surface glycoprotein: interim results from the Guinea ring vaccination cluster-randomised trial. Lancet. 2015;386: 857–866.

72. Thompson RN, Morgan OW, Jalava K. Rigorous surveillance is necessary for high confidence in end-of-outbreak declarations for Ebola and other infectious diseases. Philos Trans R Soc B. 2019;374: 20180431.

73. Lee H, Nishiura H. Sexual transmission and the probability of an end of the Ebola virus disease epidemic. J Theor Biol. 2019;471: 1–12.

74. World Health Organization. 10th Ebola outbreak in the Democratic Republic of the Congo declared over; vigilance against flare-ups and support for survivors must continue. 2020. Available: https://www.who.int/news-room/detail/25-06-2020-10th-ebola-outbreak-in-the-democratic-republic-of-the-congo-declared-over-vigilance-against-flare-ups-and-support-for-survivors-must-continue

75. Alahmadi A, Belet S, Black A, Cromer D, Flegg JA, House T, et al. Influencing public health policy with data-informed mathematical models of infectious diseases: Recent developments and new challenges. Epidemics. 2020;32: 100393.

76. Riley S, Eames K, Isham V, Mollison D, Trapman P. Five challenges for spatial epidemic models. Epidemics. 2015;10: 68–71.

77. Keeling MJ. The effects of local spatial structure on epidemiological invasions. Proc R Soc B Biol Sci. 1999;266: 859–867.

78. Mollison D. Pair approximations for spatial structures? Oberwolfach Rep. 2004;1: 2625–2627.

79. Thompson RN, Cobb RC, Gilligan CA, Cunniffe NJ. Management of invading pathogens should be informed by epidemiology rather than administrative boundaries. Ecol Modell. 2016;324: 28–32.

80. UK Government Office for National Statistics. Coronavirus-related deaths by ethnic group, England and Wales: 2 March 2020 to 10 April 2020. 2020. Available: https://www.ons.gov.uk/peoplepopulationandcommunity/birthsdeathsandmarriages/deaths/articles/coronavirusrelateddeathsbyethnicgroupenglandandwales/2march2020to10april2020

81. Zhou F, Yu T, Du R, Fan G, Liu Y, Liu Z, et al. Clinical course and risk factors for mortality of adult inpatients with COVID-19 in Wuhan, China: a retrospective cohort study. Lancet. 2020;395: 1054–1062.

82. Ng OT, Marimuthu K, Chia PY, Koh V, Chiew CJ, de Wang L, et al. SARS-CoV-2 infection among travelers returning from Wuhan, China. N Engl J Med. 2020;382: 15.

83. Thompson RN, Cunniffe NJ. The probability of detection of SARS-CoV-2 in saliva. Stat Methods Med Res. 2020;29: 1049–1050.

84. Kiss IZ, Miller JC, Simon PL. Mathematics of epidemics on networks: from exact to approximate models. Springer; 2016.

85. Pastor-Satorras R, Castellano C, Van Mieghem P, Vespignani A. Epidemic processes in complex networks. Rev Mod Phys. 2015;87: 925.

86. Álvarez LG, Webb CR, Holmes MA. A novel field-based approach to validate the use of network models for disease spread between dairy herds. Epidemiol Infect. 2011;139: 1863–1874.

87. Ball F, Sirl D, Trapman P. Analysis of a stochastic SIR epidemic on a random network incorporating household structure. Math Biosci. 2010;224: 53–73.

88. Ball F, Pellis L, Trapman P. Reproduction numbers for epidemic models with households and other social structures II: Comparisons and implications for vaccination. Math Biosci. 2016;274: 108–139.

89. Goldstein E, Paur K, Fraser C, Kenah E, Wallinga J, Lipsitch M. Reproductive numbers, epidemic spread and control in a community of households. Math Biosci. 2009;221: 11–25.

90. Pellis L, Ball F, Trapman P. Reproduction numbers for epidemic models with households and other social structures. I. Definition and calculation of R0. Math Biosci. 2012;235: 85–97.

91. Aleta A, Martin-Corral D, Piontti AP y, Ajelli M, Litvinova M, Chinazzi M, et al. Modeling the impact of social distancing testing contact tracing and household quarantine on second-wave scenarios of the COVID-19 epidemic. medrxiv. 2020.

92. Ball FG, Sirl DJ, Trapman P. Epidemics on random intersection graphs. Ann Appl Probab. 2014;24: 1081–1128.

93. Britton T, Deijfen M, Lagerås AN, Lindholm M. Epidemics on random graphs with tunable clustering. J Appl Probab. 2008;45: 743–756.

94. Haw D, Pung R, Read J, Riley S. Social networks with strong spatial embedding generate non-standard epidemic dynamics driven by higher-order clustering. bioRxiv. 2019.

95. Kiss IZ, Green DM, Kao RR. Disease contact tracing in random and clustered networks. Proc R Soc B Biol Sci. 2005;272: 1407–1414.

96. Kretzschmar ME, Rozhnova G, Bootsma M, van Boven ME, van de Wijgert J, Bonten M. Impact of delays on effectiveness of contact tracing strategies for COVID-19: a modelling study. Lancet Public Heal. 2020.

97. Ferrari MJ, Bansal S, Meyers LA, Björnstad ON. Network frailty and the geometry of herd immunity. Proc R Soc B Biol Sci. 2006;273: 2743–2748.

98. Newman MEJ. Spread of epidemic disease on networks. Phys Rev E. 2002;66: 016128.

99. Durrett R. Random Graph Dynamics. Cambridge University Press; 2006.

100. Britton T, Janson S, Martin-Löf A. Graphs with specified degree distributions, simple epidemics, and local vaccination strategies. Adv Appl Probab. 2007;39: 922–948.

101. Danon L, House TA, Read JM, Keeling MJ. Social encounter networks: Collective properties and disease transmission. J R Soc Interface. 2012;9: 2826–2833.

102. Danon L, Read JM, House TA, Vernon MC, Keeling MJ. Social encounter networks: Characterizing Great Britain. Proc R Soc B Biol Sci. 2013; 20131037.

103. World Health Organization. The First Few X (FFX) Cases and contact investigation protocol for 2019-novel coronavirus (2019-nCoV) infection. 2020. Available: https://www.who.int/publications-detail-redirect/the-first-few-x-(ffx)-cases-and-contact-investigation-protocol-for-2019-novel-coronavirus-(2019-ncov)-infection

104. Black AJ, Geard N, McCaw JM, McVernon J, Ross J V. Characterising pandemic severity and transmissibility from data collected during first few hundred studies. Epidemics. 2017;19: 61–73.

105. Cauchemez S, Ferguson NM, Wachtel C, Tegnell A, Saour G, Duncan B, et al. Closure of schools during an influenza pandemic. Lancet Infect Dis. 2009;9: 473–481.

106. Cauchemez S, Valleron A-J, Boelle P-Y, Flahault A, Ferguson NM. Estimating the impact of school closure on influenza transmission from sentinel data. Nature. 2008;452: 750–754.

107. Davies NG, Klepac P, Liu Y, Prem K, Jit M, CMMID COVID-19 Working Group, et al. Age-dependent effects in the transmission and control of COVID-19 epidemics. Nat Med. 2020.

108. Viner RM, Russell SJ, Croker H, Packer J, Ward J, Stansfield C, et al. School closure and management practices during coronavirus outbreaks including COVID-19: a rapid systematic review. Lancet Child Adolesc Heal. 2020;4: 397–404.

109. Cowling BJ, Ali ST, Ng TWY, Tsang TK, Li JCM, Fong MW, et al. Impact assessment of non-pharmaceutical interventions against coronavirus disease 2019 and influenza in Hong Kong: an observational study. Lancet Public Heal. 2020;4: 397–404.

110. Viner RM, Mytton OT, Bonell C, Melendez-Torres GJ, Ward JL, Hudson L, et al. Susceptibility to and transmission of COVID-19 amongst children and adolescents compared with adults: a systematic review and meta-analysis. medRxiv. 2020.

111. Vogel G. How Sweden wasted a "rare opportunity" to study coronavirus in schools. 2020. Available: https://www.sciencemag.org/news/2020/05/how-sweden-wasted-rare-opportunity-study-coronavirus-schools

112. McBryde ES, Trauer JM, Adekunle A, Ragonnet R, Meehan MT. Stepping out of lockdown should start with school re-openings while maintaining distancing measures. Insights from mixing matrices and mathematical models. medRxiv. 2020.

113. Teslya A, Pham TM, Godijk NG, Kretzschmar ME, Bootsma MCJ, Rozhnova G. Impact of self-imposed prevention measures and short-term government-imposed social distancing on mitigating and delaying a COVID-19 epidemic: A modelling study. PLoS Med. 2020.

114. Gilbert N. Agent-based models. 2nd Editio. Sage Publications Inc.; 2019.

115. Squazzoni F, Polhill JG, Edmonds B, Ahrweiler P, Antosz P, Scholz G, et al. Computational models that matter during a global pandemic outbreak: A call to action. J Artif Soc Soc Simul. 2020;23: 10.

116. Bryce C, Ring P, Ashby S, Wardman JK. Resilience in the face of uncertainty: early lessons from the COVID-19 pandemic. J Risk Res. 2020.

117. Kandel N, Chungong S, Omaar A, Xing J. Health security capacities in the context of COVID-19 outbreak: an analysis of International Health Regulations annual report data from 182 countries. Lancet. 2020;395: 1047–1053.

118. Cohn S, Kutalek R. Historical parallels, Ebola virus disease and cholera: Understanding community distrust and social violence with epidemics. PLoS Curr. 2016;8.

119. Kalenga OI, Moeti M, Sparrow A, Nguyen VK, Lucey D, Ghebreyesus TA. The ongoing Ebola epidemic in the Democratic Republic of Congo, 2018-2019. N Engl J Med. 2019;381: 373–383.

120. Sanicas M, Forleo E, Pozzi G, Diop D. A review of the surveillance systems of influenza in selected countries in the tropical region. Pan Afr Med J. 2014;19: 121.

121. Sankoh O, Byass P. The INDEPTH network: Filling vital gaps in global epidemiology. Int J Epidemiol. 2012;41: 579–588.

122. Wilkinson A, Ali H, Bedford J, Boonyabancha S, Connolly C, Conteh A, et al. Local response in health emergencies: key considerations for addressing the COVID-19 pandemic in informal urban settlements. Environ Urban. 2020;1: 1–20.

123. World Health Organization. Revealing the toll of COVID-19: A technical package for rapid mortality surveillance and epidemic response. 2020. Available: https://www.who.int/publications/i/item/revealing-the-toll-of-covid-19

124. Kruk ME, Pate M, Mullan Z. Introducing The Lancet Global Health Commission on High-Quality Health Systems in the SDG Era. Lancet Glob Heal. 2017;5: E480–E481.

125. Quaife M, van Zandvoort K, Gimma A, Shah K, McCreesh N, Prem K, et al. The impact of COVID-19 control measures on social contacts and transmission in Kenyan informal settlements. medRxiv. 2020.

126. Kelley M, Ferrand RA, Muraya K, Chigudu S, Molyneux S, Pai M, et al. An appeal for practical social justice in the COVID-19 global response in low-income and middle-income countries. Lancet Glob Heal. 2020;395: 1047–1053.

127. University of Oxford Blavatnik School of Government. Coronavirus government response tracker.

2020.

128. Dong E. An interactive web-based dashboard to track COVID-19 in real time. Lancet Infect Dis. 2020;20: 533–534.

129. Kenah E. Contact intervals, survival analysis of epidemic data, and estimation of R0. Biostatistics. 2010;12: 548–566.

130. Oliver N, Lepri B, Sterly H, Lambiotte R, Delataille S, De Nadai M, et al. Mobile phone data for informing public health actions across the COVID-19 pandemic life cycle. Sci Adv. 2020; eabc0764.

131. Sun K, Chen J, Viboud C. Early epidemiological analysis of the coronavirus disease 2019 outbreak based on crowdsourced data: a population-level observational study. Lancet Digit Heal. 2020;2: e201–e208.

132. Leclerc Q, Nightingale ES, Abbott S, CMMID nCoV Working Group, Jombart T. Analysis of temporal trends in potential COVID-19 cases reported through NHS Pathways England. medRxiv. 2020.

133. Prem K, Cook AR, Jit M. Projecting social contact matrices in 152 countries using contact surveys and demographic data. PLoS Comput Biol. 2017;13: e1005697.

134. Klepac P, Kissler S, Gog J. Contagion! The BBC Four Pandemic – The model behind the documentary. Epidemics. 2018;24: 49–59.

135. Ferguson NM, Laydon D, Nedjati-Gilani G, Imai N, Ainslie K, Baguelin M, et al. Report 9: Impact of non-pharmaceutical interventions (NPIs) to reduce COVID-19 mortality and healthcare demand. 2020. Available: www.imperial.ac.uk/mrc-global-infectious-disease-analysis/covid-19/report-9-impact-of-npis-on-covid-19/

136. Dehning J, Zierenberg J, Spitzner EP, Wibral M, Neto JP, Wilczek M, et al. Inferring change points in the spread of COVID-19 reveals the effectiveness of interventions. Science (80- ). 2020; eabb9789.

137. Gilbert M, Pullano G, Pinotti F, Valdano E, Poletto C, Boëlle PY, et al. Preparedness and vulnerability of African countries against importations of COVID-19: a modelling study. Lancet. 2020;395.

138. Kifer D, Bugada D, Villar-Garcia J, Gudelj I, Menni C, Sudre CH, et al. Effects of environmental factors on severity and mortality of COVID-19. medRxiv. 2020.

139. Al-Rousan N, Al-Najjar H. The correlation between the spread of COVID-19 infections and weather variables in 30 Chinese provinces and the impact of Chinese government mitigation plans. Eur Rev Med Pharmacol Sci. 2020;24: 4565–4571.

140. Liu J, Zhou J, Yao J, Zhang X, Li L, Xu X, et al. Impact of meteorological factors on the COVID-19 transmission: A multi-city study in China. Sci Total Environ. 2020;726: 138513.

141. Qi H, Xiao S, Shi R, Ward MP, Chen Y, Tu W, et al. COVID-19 transmission in Mainland China is associated with temperature and humidity: A time-series analysis. Sci Total Environ. 2020;19: 138778.

142. Şahin M. Impact of weather on COVID-19 pandemic in Turkey. Sci Total Environ. 2020;728: 138810.

143. Menebo MM. Temperature and precipitation associate with Covid-19 new daily cases: A correlation study between weather and Covid-19 pandemic in Oslo, Norway. Sci Total Environ. 2020;737: 139659.

144. Tosepu R, Gunawan J, Effendy DS, Ahmad LOAI, Lestari H, Bahar H, et al. Correlation between weather and Covid-19 pandemic in Jakarta, Indonesia. Sci Total Environ. 2020;4: 138436.

145. Araujo MB, Naimi B. Spread of SARS-CoV-2 Coronavirus likely to be constrained by climate. medRxiv. 2020.

146. Coro G. A global-scale ecological niche model to predict SARS-CoV-2 coronavirus infection rate. Ecol Modell. 2020;431: 109187.

147. Carlson CJ, Chipperfield JD, Benito BM, Telford RJ, O'Hara RB. Species distribution models are inappropriate for COVID-19. Nat Ecol Evol. 2020;4: 770–771.

148. European Centre for Medium-Range Weather Forecasts. The ERA5 dataset. [cited 15 Jul 2020]. Available: https://www.ecmwf.int/en/forecasts/datasets/reanalysis-datasets/era5

149. Hurrell JW, Holland MM, Gent PR, Ghan S, Kay JE, Kushner PJ, et al. The community earth system model: A framework for collaborative research. Bull Am Meteorol Soc. 2013;94: 1339–60.

150. Baker RE, Yang W, Vecchi GA, Metcalf CJE, Grenfell BT. Susceptible supply limits the role of

climate in the early SARS-CoV-2 pandemic. Science (80- ). 2020;369: 315–319.

151. Mecenas P, Bastos R, Vallinoto A, Normando D. Effects of temperature and humidity on the spread of COVID-19: A systematic review. medRxiv. 2020.

152. Li Q, Guan X, Wu P, Wang X, Zhou L, Tong Y, et al. Early transmission dynamics in Wuhan, China, of novel coronavirus–infected pneumonia. N Engl J Med. 2020.

153. De Angelis D, Presanis AM, Birrell PJ, Tomba GS, House T. Four key challenges in infectious disease modelling using data from multiple sources. Epidemics. 2015;10: 83–87.

154. Anderson RM, Heesterbeek H, Klinkenberg D, Hollingsworth TD. How will country-based mitigation measures influence the course of the COVID-19 epidemic? Lancet. 2020;395: 931–934.

155. Dawid AP. Prequential data analysis. Current Issues in Statistical Inference: Essays in Honor of D Basu. 1992. pp. 113–126.

156. Reich NG, McGowan CJ, Yamana TK, Tushar A, Ray EL, Osthus D, et al. Accuracy of real-time multi-model ensemble forecasts for seasonal influenza in the U.S. PLoS Comput Biol. 2019;15: e1007486.

157. Shea BK, Runge MC, Pannell D, Probert WJM, Li SL, Tildesley M, et al. Harnessing multiple models for outbreak management. Science (80- ). 2020;368: 577–579.

158. Thompson RN, Gilligan CA, Cunniffe NJ. Control fast or control smart: When should invading pathogens be controlled? PLoS Comput Biol. 2018;14.

159. Zimmer C, Leuba SI, Cohen T, Yaesoubi R. Accurate quantification of uncertainty in epidemic parameter estimates and predictions using stochastic compartmental models. Stat Methods Med Res. 2019; 3591–3608.

160. Chowell G. Fitting dynamic models to epidemic outbreaks with quantified uncertainty: A primer for parameter uncertainty, identifiability, and forecasts. Infect Dis Model. 2017;2: 379–398.

161. Shearer FM, Moss R, McVernon J, Ross J V, McCaw JM. Infectious disease pandemic planning and response: Incorporating decision analysis. PLoS Med. 2020;17: e1003018.

162. Lipsitch M, Finelli L, Heffernan RT, Leung GM, Redd SC. Improving the evidence base for decision making during a pandemic: The example of 2009 influenza A/H1N1. Biosecurity Bioterrorism Biodefense Strateg Pract Sci. 2011;9: 89–115.

163. Lipsitch M, Donnelly CA, Fraser C, Blake IM, Cori A, Dorigatti I, et al. Potential biases in estimating absolute and relative case-fatality risks during outbreaks. PLoS Negl Trop Dis. 2015;9: e0003846.

164. Verity R, Okell LC, Dorigatti I, Winskill P, Whittaker C, Imai N, et al. Estimates of the severity of coronavirus disease 2019: a model-based analysis. Lancet Infect Dis. 2020;20: 669–677.

**Acknowledgements**

The authors would like to thank the Isaac Newton Institute for Mathematical Sciences, Cambridge (www.newton.ac.uk), for support during the virtual "Infectious Dynamics of Pandemics" programme where work on this paper was begun. This work was undertaken in part as a contribution to the "Rapid Assistance in Modelling the Pandemic" initiative coordinated by the Royal Society. Thanks to Sam Abbott for helpful comments about the original manuscript.

**Funding**

This work was supported by the Isaac Newton Institute (EPSRC grant number EP/R014604/1). RNT thanks Christ Church (Oxford) for funding via a Junior Research Fellowship. RNT and SF acknowledge support from the Wellcome Trust (210758/Z/18/Z). LHKC acknowledges support from the BBSRC (BB/R009236/1). BA is supported by the Natural Environment Research Council (NE/N014979/1). CAD and KVP thank the UK MRC and DFID for centre funding (MR/R015600/1). CAD also thanks the UK NIHR (National Institute for Health Research) HPRU (Health Protection Research Unit). RME acknowledges HDR UK (MR/S003975/1) and the UK MRC (MC_PC 19065). HH and MEK acknowledge support from the Netherlands Organisation for Health Research and Development (ZonMw; grant 10430022010001). TH acknowledges support from the Royal Society (INF\R2\180067) and the Alan Turing Institute for Data Science and Artificial Intelligence. MK and MJT acknowledge support from the UK MRC (MR/V009761/1). IZK acknowledges support from the Leverhulme Trust (RPG-2017-370). MEK acknowledges support from the Netherlands Organisation for Health Research and Development (ZonMw; grant 91216062). JCM acknowledges startup funding from La Trobe University. CABP acknowledges funding of the NTD Modelling Consortium by the Bill and Melinda Gates Foundation (OPP1184344). LP acknowledges support from the Wellcome Trust and the Royal Society (202562/Z/16/Z). JRCP acknowledges support from the South African Centre for Epidemiological Modelling and Analysis (SACEMA), a Department of Science and Innovation-National Research Foundation Centre of Excellence hosted at Stellenbosch University. CJS acknowledges support from CNPq and FAPERJ. PT acknowledges support from Vetenskapsrådet Swedish Research Council (2016-04566).

**Ethics statement**

The authors declare that no ethical concerns exist.

**Data accessibility statement**

Data sharing is not applicable to this manuscript as no new data were created or analysed in this study.

**Competing interests statement**

The authors declare that no competing interests exist.

**Authors' contributions**

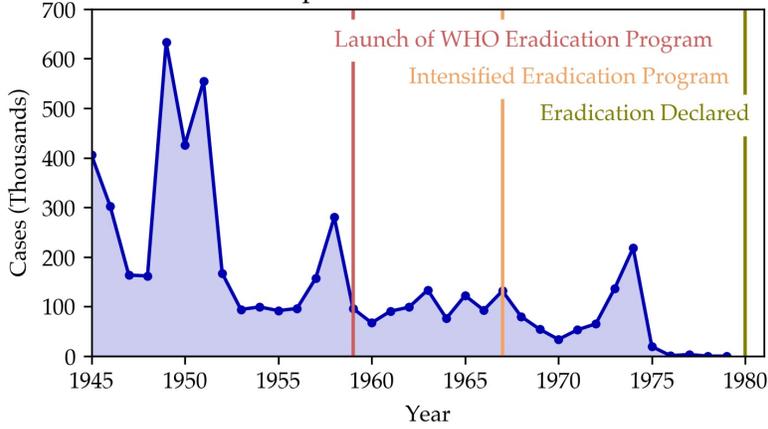
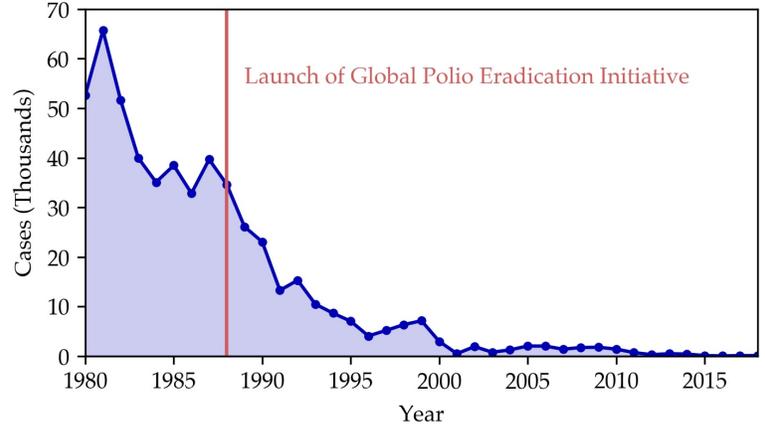
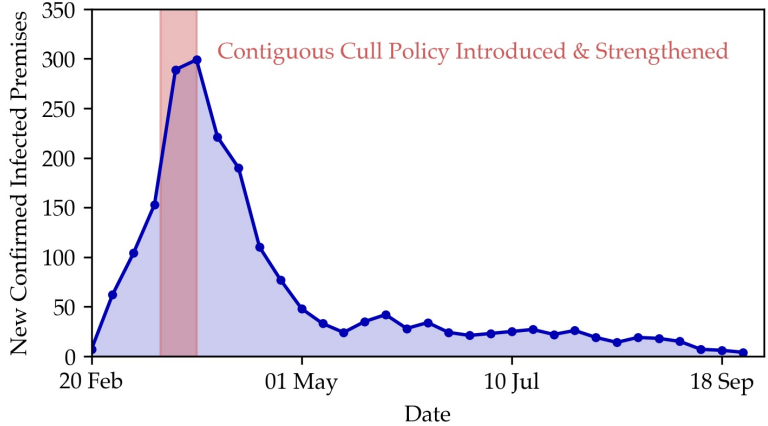
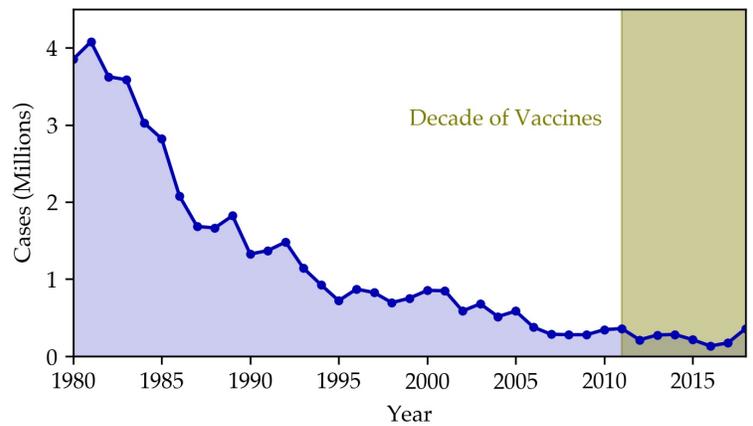
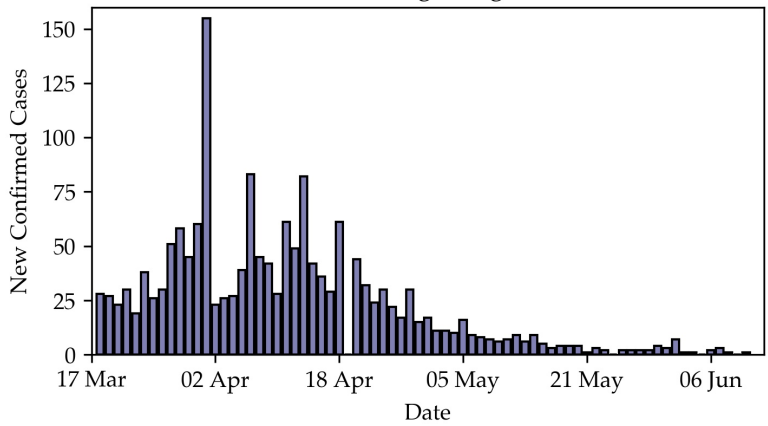
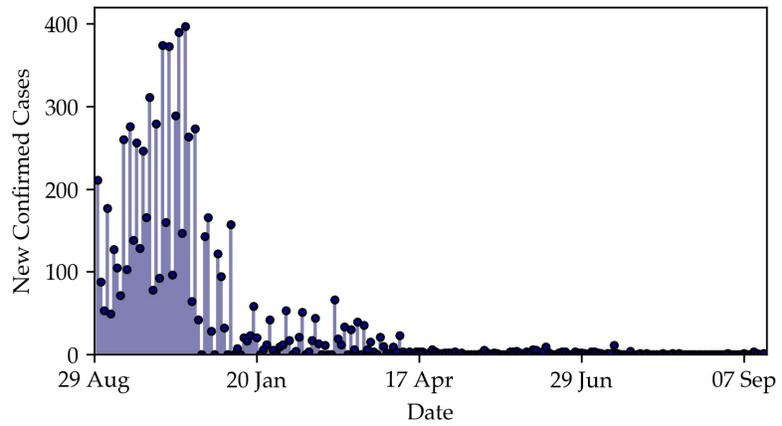